 \documentclass[preprint2]{aastex}
\usepackage{graphics,epsfig,longtable,lscape,txfonts,color}
\usepackage[english]{babel}

\newcommand{\myemail}{hstiele@mx.nthu.edu.tw}

\newcommand{\mr}{\mathrm}
\newcommand{\nh}{\hbox{$N_{\mr H}$}}
\newcommand{\hcm}[1]{$\times 10^{#1}$ cm$^{-2}$}
\def\subsun{\mbox{$_{\odot}$}}

\newcommand{\ergs}[1]{$\times10^{#1}$~\hbox{erg~s$^{-1}$}}
\newcommand{\oergs}[1]{$10^{#1}$~erg~s$^{-1}$}

\def\ie{i.\,e.}                                      
\def\eg{e.\,g.}                                      

\def\xmm{\textit{XMM-Newton}}

\def\gx339{GX\,339-4}
\def\h1743{H\,1743-322}

\def\max15{MAXI\,J1535--571}
\def\max18{MAXI\,J1820+070}
\def\xu12{XMMU 122939.7+075333}
\def\mnras{MNRAS}
\def\aap{A\&A}

\def\apjl{ApJ}
\def\aj{AJ}
\def\apj{ApJ}

\def\pasp{PASP}
\def\araa{Annual Rev. of A\&A}

\def\nat{Nature}

\shorttitle{2018 outburst of \max18}
\shortauthors{Stiele, Kong}

\begin{document}


\title{Peculiar outbursts of an ultra luminous source likely signs of an aperiodic disc-wind}

\author{H.\ Stiele}
\affil{Institute of Astronomy, National Tsing Hua University, No.~101 Sect.~2 Kuang-Fu Road,  30013, Hsinchu, Taiwan}
\email{\myemail}
\and

\author{A.\ K.\ H.\ Kong}
\affil{Institute of Astronomy, National Tsing Hua University, No.~101 Sect.~2 Kuang-Fu Road,  30013, Hsinchu, Taiwan}





\begin{abstract}
The metal rich globular cluster RZ 2109 in the massive Virgo elliptical galaxy NGC 4472 (M49) harbours the ultra luminous X-ray source XMMU 122939.7+075333. Previous studies showed that this source varies between bright and faint phases on timescales of just a few hours. Here, we report the discovery of two peculiar X-ray bursting events that last for about 8 and 3.5 hours separated by about 3 days. Such a recurring X-ray burst-like behaviour has never been observed before. We argue that type-I X-ray bursts or super bursts as well as outburst scenarios requiring a young stellar object are highly unlikely explanations for the observed light curve, leaving an aperiodic disc wind scenario driven by hyper-Eddington accretion as a viable explanation for this new type of X-ray flaring activities. 
\end{abstract}


\section{Introduction}
The ultra luminous X-ray source (ULX) XMMU 122939.7+075333 resides in the spectrally confirmed, metal rich globular cluster RZ 2109 in the massive Virgo elliptical galaxy NGC 4472 (M49) \citep{2001AJ....121..210R}, which is at a distance of $17.14\pm0.71$ Mpc \citep{2008ApJ...676..184T}. Most times it is observed at a soft X-ray luminosity $L_{\mr{x};\, 0.2 - 10\ \mr{keV}}\sim3$\ergs{39}, apart from one observation in 2010, where it was observed at a lower luminosity \citep[$L_{\mr{x}}\sim1$\ergs{38};][]{2015MNRAS.447.1460J}. Previous studies found that the source varies between bright and faint phases on timescales of just a few hours \citep{2007Natur.445..183M,2008MNRAS.386.2075S}. This variability has been addressed to changes in the absorption column density and lead to the classification of \xu12\ as a black hole system. Investigations of the optical emission (especially of the O\textsc{iii} emission lines) associated with \xu12\ prefer a stellar mass black hole primary in this source over an intermediate mass black hole \citep{2008ApJ...683L.139Z,2011ApJ...739...95S}. The observed variability rules out a sample of neutron stars as explanation for the huge luminosity of \xu12 \citep{2007Natur.445..183M,2008MNRAS.386.2075S}. However, a single neutron star accreting at super-Eddington rate can be an alternative, as more recent studies detected pulsed emission from other ULXs, which shows that these systems can harbour a single neutron star  \citep{2014Natur.514..202B,2017arXiv170200966T}.

\section[]{Observation and data analysis}
\label{Sec:obs}
We analysed three \xmm\ observations of \xu12\ taken in January 2016 with the EPIC cameras in imaging mode. Details on the observations are given in table \ref{Tab:data}. We filtered and extracted the pn event file, using standard SAS (version 14.0.0) tools, paying particular attention to extract the list of photons not randomized in time. We extracted light curves in the 10 -- 12 keV range (binning 100s) to identify periods of increased background radiation (count rate $>0.4$cts/s) that we then excluded from our study ($\sim20 - 25$ ks in each observation, see exposure column in table \ref{Tab:data}). From the remaining data (not affected by background flares) we extracted light curves and energy spectra using a circle with a radius of 10\arcsec\ placed at the nominal position of the source, and corresponding background spectra from a source free region on the same chip, redistribution matrices, and ancillary response files. We included single and double events (PATTERN$\le$4) in our study.
 
The light curves in the 0.2 -- 10 keV band (binning 1ks) were corrected for different effects, such as vignetting, bad pixels, dead time, including background subtraction using the \textsc{sas} task \texttt{epiclccorr}. From these light curves we selected periods when \xu12\ was bright (count rate $\ge0.055$cts/s) or faint (count rate $\le0.015$cts/s) and extracted the corresponding spectra. 

\begin{table}
\caption{Details of \xmm\ observations}
\begin{center}
\begin{tabular}{lrrrr}
\hline\noalign{\smallskip}
 \multicolumn{1}{c}{\#} & \multicolumn{1}{c}{Obs.~id.} & \multicolumn{1}{c}{Date}  &  \multicolumn{2}{c}{Exposure [ks]}  \\
& &  &  \multicolumn{1}{c}{net}  &  \multicolumn{1}{c}{gti$^{\dagger}$}   \\
 \hline\noalign{\smallskip}
1 & 0761630101  & 2016-01-05  & 100.7 & 79.0 \\
\noalign{\smallskip}
2 & 0761630201 & 2016-01-07  & 98.3 & 72.1 \\
\noalign{\smallskip}
3 & 0761630301 & 2016-01-09  & 100.1 & 74.4 \\
\noalign{\smallskip}
\hline\noalign{\smallskip} 
\end{tabular} 
\end{center}
Notes: \\
$^{\dagger}$: applying good time intervals\\
\label{Tab:data}
\end{table}
 
\section[]{Results}
\label{Sec:res}

\subsection{Light curves}
\begin{figure}
\resizebox{\hsize}{!}{\includegraphics[clip,angle=0]{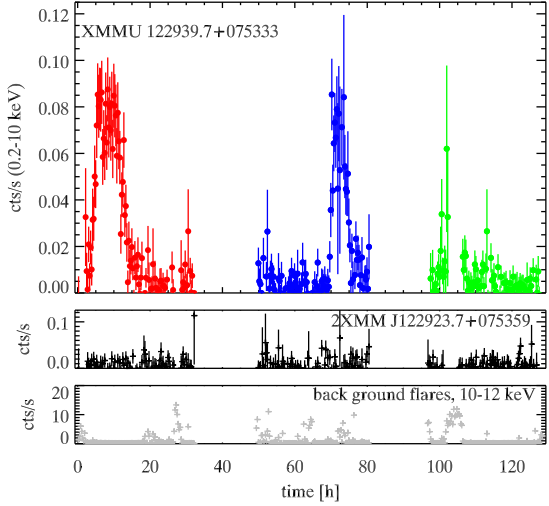}}
\caption{Light curves of all three observations in the 0.2 -- 10 keV band, binning 1ks. $T=0$ corresponds to the beginning of the first observation (2016-01-05 09:14:09.746 UTC). The two lower panels show the 0.2 -- 10 keV band relative light curve ($r_{\mr{cts}}-<r_{\mr{cts}}>$) of the quasar 2XMM J122923.7+075359 and an uncorrected light curve extracted from all events detected in the 10 -- 12 keV range, to check for background flares, respectively. The gap in the third observation is due to a background flare (the increase in count rate before the gap is most likely related to incomplete removal of the background flare).}
\label{Fig:lc}
\end{figure}

An \xmm\ EPIC/pn light curve of \xu12\ taken in January 2016 provides a good (although not complete) coverage of 5.7 days (Fig.~\ref{Fig:lc}), consisting of three observations with two gaps of $\sim$17.5 h between the observations. What is really astonishing about this light curve (Fig.~\ref{Fig:lc}), is that we observe two flares within about 3 days. These two flares have a duration of 7.78 h and 3.42 h, respectively, and reach averaged luminosities of $L_{\mr{x}}\sim3.22$\ergs{38} and $L_{\mr{x}}\sim1.98$\ergs{38}, respectively. These flares are not related to flaring in the background rate, as can be seen by comparing the light curve of \xu12\ to the uncorrected light curve extracted from all events detected in the 10 -- 12 keV range in the total exposure, which is also shown in Fig.~\ref{Fig:lc}. In addition, we also show the relative light curve (count rate minus mean count rate: $r_{\mr{cts}}-<r_{\mr{cts}}>$, where $<r_{\mr{cts}}>=$ 0.069, 0.074, and 0.078 cts/s for the three observations respectively) of the quasar 2XMM J122923.7+075359 \citep{2011MNRAS.410..860A,2016A&A...593A..55V} in the 0.2 -- 10 keV band, and this light curve does not show any flaring. This gives further evidence, that the variability observed in the corrected light curve of \xu12\ is intrinsic to this source and not related to any background flares present in the raw data. Based on the currently available data we can draw no firm conclusion if the observed brightening is periodic or not. If the spacing between the two observed flares would be the period ($\sim$70 h), the onset of the next flare would take place $\sim$12.5 h after the end of the third observation, consistent with the persistent emission throughout the third observation. The fact that the duration and averaged luminosity of the second flare are only about halve of that of the first observation, questions the periodic nature. 

\begin{figure}
\resizebox{\hsize}{!}{\includegraphics[clip,angle=0]{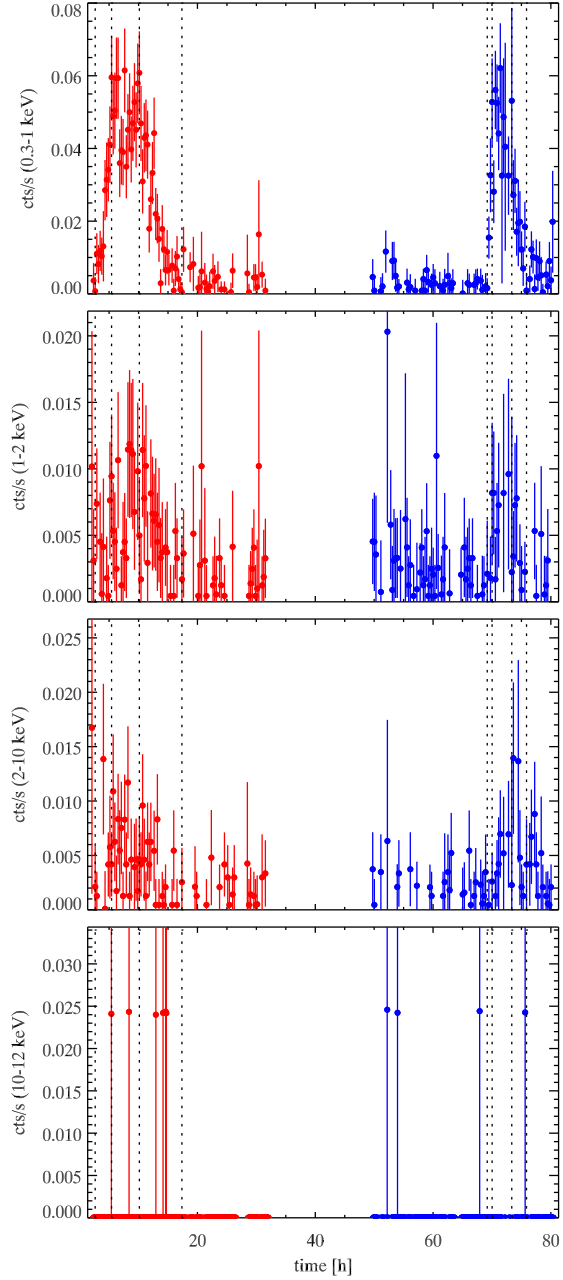}}
\caption{Light curves of the first two observations (that show flares) in three different energy bands. Dotted lines indicate the rise and decay of the flare.}
\label{Fig:lc_edep}
\end{figure}

Energy resolved light curves in four different bands for the first two observations are shown in Fig.\ \ref{Fig:lc_edep}. In both observations the flares are clearly visible in the softest energy band (0.3 -- 1 keV). The rising part of the 0.3 -- 1 keV light curves can be described with a power-law while an exponential decay model fits better for the decay part (see Table~\ref{Tab:lc_soft}). On the other hand, the 1 -- 2 and 2 -- 10 keV light curves are consistent with a steady source (Table~\ref{Tab:lc_soft}). The light curve in the 10 -- 12 keV band does not contain any source photons and only shows statistical noise. If we assume that the duration of the rise corresponds to the light-crossing timescale, the compact object should have about 1.2 to 7.3 $\times 10^8$ M\subsun\ \citep{2011MNRAS.412.1389N}, implying a supermassive black hole. We can rule out that the source is a supermassive black hole, as the X-ray source is associated with a globular cluster.   

\begin{table}
\caption{Results of soft band (0.3 -- 1 keV) light curve fit.}
\begin{center}
\begin{tabular}{ll}
\hline\noalign{\smallskip}
 \multicolumn{1}{c}{obs.\ 1} & \multicolumn{1}{c}{obs.\ 2}   \\
 \hline\noalign{\smallskip}
  \multicolumn{2}{c}{rise; power law}\\ 
$\Gamma=1.38_{-0.18}^{+0.20}$  &  $\Gamma=1.63\pm0.37$\\
\noalign{\smallskip}
$\chi^2/dof: 6.19/14$  &  $\chi^2/dof: 0.27/2$\\
 \hline\noalign{\smallskip}
  \multicolumn{2}{c}{decay; power law}\\ 
$\Gamma=-0.45\pm0.04$ & $\Gamma=-0.53_{-0.09}^{+0.11}$  \\
\noalign{\smallskip}
$\chi^2/dof: 52.85/19$  &  $\chi^2/dof: 8.50/8$\\
 \hline\noalign{\smallskip}
  \multicolumn{2}{c}{decay; exponential}\\ 
$\alpha=(-1.2\pm0.1)\times10^{-4}$ & $\alpha=(-2.1\pm0.6)\times10^{-4}$ \\
\noalign{\smallskip}
$\chi^2/dof: 25.34/19$  &  $\chi^2/dof: 5.84/8$\\
\noalign{\smallskip}
\hline\noalign{\smallskip} 
\end{tabular} 
\end{center}
\label{Tab:lc_soft}
\end{table}

\subsection{Spectral properties}
\label{Sec:spec}
We fit the averaged energy spectra of the flares and persistent emission within \textsc{isis} \citep[V.~1.6.2;][]{2000ASPC..216..591H} in the 0.3 -- 10 keV range. We grouped the data to contain at least 20 counts per bin. The flare spectra of the first two observations can be described with a model consisting of thermal disc emission (\texttt{diskbb}) and emission of an O\,\textsc{viii} line, modelled using a \texttt{gaussian}. In case of the first observation this model gives a statistically unacceptable fit with $\chi^2/dof = 79.5/30$ and strong residuals at energies above 1 keV are observed, which shows that Comptonised emission is also present. To model this Comptonised component we include the \texttt{simpl} model \citep{2009PASP..121.1279S} in our fit of the flare spectrum of the first observation, which results in a statistically acceptable fit with $\chi^2/dof = 13.4/27$. In case of the second observation no component to account for Comptonized emission is needed, as the flare spectrum does not contain photons above $\sim$2 keV. The spectral parameters obtained for both observations are given in table \ref{Tab:spec}. The parameters of the thermal component and of the O\,\textsc{viii} line are consistent within errors between the first and second observation, apart from the \texttt{diskbb} normalisation which is much smaller in the second observation. The accretion disc temperature agrees well with the values obtained in previous studies \citep{2007Natur.445..183M,2015MNRAS.447.1460J}. 

The spectra of the persistent emission of all observations can be fitted by thermal disc emission. In case of the first observation we need to add emission of an O\,\textsc{viii} line. Alternatively, the spectra of the persistent emission of the second and third observation can be well fitted with Comptonised emission (power law). All spectral parameters can be found in table \ref{Tab:spec}. While the spectrum of persistent emission of the first observation does not contain photons above $\sim$2 keV, photons at higher energies are present in the spectra of the second and third observation.

Our spectral study confirms previous findings that the foreground absorption in the persistent emission is higher than during the flare \citep{2007Natur.445..183M,2015MNRAS.447.1460J}. However, we cannot confirm that the models used to fit the spectra of the flares give acceptable fits of the persistent emission spectra, as these models contain components that are not required to obtain acceptable fits of the persistent emission spectra (the component to fit the Componized emission in case of the first observation, and the component to fit the O\,\textsc{viii} line in case of the second observation).

The fast-rise-exponential-decay outburst profile is similar to that of type-I X-ray bursts of neutron star X-ray binaries. To model the possible emission from the X-ray burst, we fit the flare spectra by using the best-fit persistent spectra and an additional blackbody component. In case of the first observation, we also need to include a Comptonised component and allow for a free normalisation of the \texttt{gaussian} that fits the O\,\textsc{viii} line, \ie\ the strength of the O\,\textsc{viii} line can vary between flare and persistent emission spectra. To obtain a statistically acceptable fit we need to assume that the Comptonised emission is caused by up scattering of photons of the added blackbody component. If instead the Comptonised photons are assumed to be due to up scattering of the disk blackbody component, the fits are statistically poor ($\chi^2/dof = 39.6/29$) and strong residuals at energies above 1 keV are present. In the second observation we need to include the O\,\textsc{viii} line, which is not needed to fit the persistent emission spectra of this observation. The obtained spectral parameters are given in table \ref{Tab:spec}. The absorption of the added components is above the Galactic foreground absorption \citep[\nh$=1.6$\hcm{20};][]{1990ARA&A..28..215D}, but much lower than the one needed to fit the persistent emission spectra. For the first observation the absorption agrees with the value found for the disk blackbody plus Comptonised emission model (for the second observation  the absorption is fixed in both cases). The temperature of the blackbody component does not change between both observations, and it is close to the value of the accretion disk temperature observed during the flare. The emission area of the blackbody radiation seems to be smaller in the second observation.

\begin{table*}
\caption{Best fit parameters for the averaged \xmm/pn spectra of XMMU 122939.7+075333} 
\begin{center}
\begin{tabular}{llll}
\hline\noalign{\smallskip}
\multicolumn{4}{c}{flare spectra}\\
\hline\noalign{\smallskip}
 \multicolumn{1}{c}{parameter} & \multicolumn{1}{c}{obs.\ 1} & \multicolumn{1}{c}{obs.\ 2}&\multicolumn{1}{c}{obs.\ 3} \\
 \hline\noalign{\smallskip}
n$_{\mr{H}} 10^{20}$ cm$^{-2}$ &  $1.86_{-1.86}^{+ 4.01} $ &  $1.86^{\dagger} $  &  -- \\         
\smallskip
T$_{\mr{in}}$ keV &  $0.12_{-0.03}^{+0.02} $  &  $0.18_{-0.03}^{+0.02} $ & --\\             
\smallskip
R$_{\mr{in}}^{\ddagger}$ km&  $11584.41^{+15571.07}_{-4650.64} $ &  $4373.22^{+1950.32}_{-1415.82} $ &   -- \\               
\smallskip
$\Gamma$ &  $1.39_{-0.29}^{+0.32} $  &  --&  --\\            
\smallskip
FracSctr  &  $0.13_{-0.05}^{+0.17} $ &  -- &  --\\                 
\smallskip
E$_{\mr{OVIII}}$ keV & $0.64_{-0.02}^{+0.03} $  & $0.65\pm0.05 $ & -- \\         
\smallskip
EW$_{\mr{OVIII}}$ eV & $55.72_{-33.55}^{+35.28} $  & $51.43_{-51.43}^{+40.50} $& -- \\         
\smallskip
gaus.\ norm $\times10^{-5}$& $0.97_{-0.34}^{+0.71} $   &$1.18_{-0.61}^{+0.86} $ & -- \\           
\smallskip
$L_{\mr{x}}$\ergs{38} & 3.22& 1.98& --\\ 
\smallskip
$\chi^2/dof$ &  13.4/27& 3.8/3& -- \\
\smallskip
duration h & 7.78 & 3.42 & --\\
\hline\noalign{\smallskip} 
\multicolumn{4}{c}{persistent spectra}\\
\hline\noalign{\smallskip}
n$_{\mr{H}} 10^{20}$ cm$^{-2}$ &  $14.14_{-14.14}^{+60.02} $ &  $<3.85$ (dbb) /  $3.72_{-3.72}^{+11.99} $ (pl)& $<4.55$ (dbb) / $4.42_{-4.42}^{+12.24} $ (pl)\\         
\smallskip
T$_{\mr{in}}$ keV &  $ 0.31_{-0.13}^{+0.24} $ &  $ 0.29\pm0.07 $ & $ 0.59_{-0.13}^{+0.18} $  \\             
\smallskip
R$_{\mr{in}}^{\ddagger}$ km$^2$ &  $371.59^{ +4736.13}_{-298.11} $ &  $ 390.85^{+2730.09}_{-158.50} $ &  $93.88^{+2354.24}_{-41.92} $  \\               
\smallskip
& & ------ & ------ \\
$\Gamma$ &  --  & $2.77_{-0.53}^{+1.05} $ &  $2.04_{-0.37}^{+0.57} $ \\            
\smallskip
pl.\ norm $\times10^{-6}$&  --  &   $1.87_{-0.48}^{+0.72} $ &  $1.88_{-0.48}^{+0.88} $\\          
\smallskip
E$_{\mr{OVIII}}$ keV & $0.62^{\dagger} $ &  --  & -- \\         
\smallskip
EW$_{\mr{OVIII}}$ eV & $30.00_{-30.00}^{+50.81} $ &  -- & -- \\         
\smallskip
gaus.\ norm $\times10^{-6}$& $3.30_{-2.22}^{+37.23E6} $ &  --  & -- \\           
\smallskip
$L_{\mr{x}}$\ergs{37} & 1.59&  1.58 (dbb) / 2.06 (pl)& 1.76 (dbb) / 2.57 (pl)\\ 
\smallskip
$\chi^2/dof$ &  3.1/3 &  10.1/8 (dbb) / 7.6/8 (pl)  & 7.2/9 (dbb) / 6.0/9 (pl) \\
\hline\noalign{\smallskip} 
\multicolumn{4}{c}{flare spectra fitted with additional blackbody component (and further components as needed)}\\
\hline\noalign{\smallskip} 
n$_{\mr{H}} 10^{20}$ cm$^{-2}$ &   $<3.20$ &  $1.86^{\dagger} $  &  -- \\         
\smallskip
kT$_{\mr{bb}}$ keV &    $0.10\pm0.01 $  &  $0.12\pm0.01 $  (dbb/pl)& --\\             
\smallskip
R$_{\mr{bb}}^{\ddagger}$ km &    $16631.98^{+6701.05}_{-2846.44} $ &  $10565.81^{+3254.98}_{-2385.38} $ (dbb) /  $10340.98^{+3148.48}_{-2258.48} $ (pl)   & --\\               
\smallskip
$\Gamma$  &  $1.13_{-0.03}^{+0.30} $  &  -- & --\\            
\smallskip
FracSctr  &    $0.28_{-0.14}^{+0.07} $ &  --  & --\\                 
\smallskip
E$_{\mr{OVIII}}$ keV &  -- &  $0.69_{-0.03}^{+0.01} $  (dbb/pl) & --\\         
\smallskip
EW$_{\mr{OVIII}}$ eV &  -- &  $<89.6 $ (dbb) / $<90.3 $ (pl)& --\\         
\smallskip
gaus.\ norm $\times10^{-5}$& $1.30\pm0.42$ &  $0.75_{-0.36}^{+0.37} $ (dbb) / $0.76\pm0.36 $ (pl) & --\\           
\smallskip
$L_{\mr{x}}$\ergs{38} &  3.32 & 1.95 (dbb) /  1.99 (pl)& --\\ 
\smallskip
$\chi^2/dof$ &   16.5/29 & 4.6/3 (dbb) / 4.5/3 (pl)& --\\
\smallskip
duration h & 7.78 & 3.42&  --\\
\hline\noalign{\smallskip}
\end{tabular} 
\end{center}
Notes: \\
$^{\dagger}$: fixed\\
$^{\ddagger}$: assuming a distance of $17.14\pm0.71$ Mpc and no corrections included\\
See Sect.~\ref{Sec:spec} for details on the models that have been used to fit flare / persistent emission of different observations.
\label{Tab:spec}
\end{table*}

\section[]{Discussion}
\label{Sec:dis}
Interpreting the bright phases as type-I X-ray bursts is hampered by the fact that the time scales are highly inconsistent, as type-I X-ray bursts last for at most a few minutes and not several hours. Regarding the duration of the flares, super bursts which have been observed in a few neutron star X-ray binaries \citep{2004NuPhS.132..466K}, would seem to be a better explanation. However, the observed recurrence time would be incredible short for a super burst \citep{2006A&A...455.1031K} and the temperature obtained from energy spectra would be much too low \citep{2003astro.ph..1544S,2004NuPhS.132..466K}. Furthermore, the radius of the thermal emission area is $\sim10^3-10^4$ km, much larger than a neutron star. We can also exclude tidal disruption events as a possible explanation, as the slope of the power law fitted to the decay from bright to faint phase ($\sim-1/2$; see Table~\ref{Tab:lc_soft}) is much flatter than the $-5/3$ decay characteristic for tidal disruption events \citep{1988Natur.333..523R,1989IAUS..136..543P,1989ApJ...346L..13E}. In case of partial disruption, which could help to explain why we would observe two disruption events within $\sim3$ days, one would expect to observe an even steeper slope \citep{2013ApJ...767...25G}. Pervious optical studies of \xu12\ have already considered tidal disruption events an unlikely explanation for the observed variability of this source \citep{2008ApJ...683L.139Z,2011ApJ...739...95S}.   

In case of a neutron star primary, magnetar outbursts are a possible explanation for the observed light curve. However, we do not expect to find a magnetar in the old population of a globular cluster.
 
Outbursts lasting about 10 h, reaching a peak luminosity of \oergs{38}, but with a much shorter period of about 14--15 h, have been observed for CXOU J004732.0--251722.1 in the starburst galaxy NGC 253 and this source has been classified as a Wolf-Rayet X-ray binary \citep{2014MNRAS.439.3064M}. This does not seem to be a viable explanation for \xu12\ as again we do not expect to find a Wolf-Rayet star in the old population of a globular cluster.
 
Ultra luminous flares from globular clusters or ultra compact dwarf companions of nearby parent elliptical galaxies, reaching luminosities of $10^{39}$--\oergs{41}, have been observed \citep{2005ApJ...624L..17S,2016Natur.538..356I}. However, the rise times of these flares were less than one minute, and the flares then decayed over about an hour. It is much shorter time scales than the ones observed in \xu12.

Giant outburst of Cyg X-1 have been observed in the hard X-rays (15 -- 300 keV). These outbursts show a large variety in duration from less than one hour to more than 7 hours, and they do not show any periodicity. They have been suggested to be related to some rare eruptive phenomenon in the donor wind ejection \citep{2003ApJ...596.1113G}. This also does not seem to be a viable explanation for the light curve of \xu12, as we do not expect to find a young, high-mass donor star in the old population of a globular cluster.

The presence of the O\textsc{iii} emission lines, requires that there is some O rich gas in the vicinity of the compact object, which is irradiated by the X-ray emission of the compact object. If this gas fuels the outbursts it must be replenished quite regularly. If it obscures the emission of the compact object, it must be diluted form time to time to allow for the bright phases. This dilution can be caused \eg\ by the irradiation of the gas by emission form the compact object. In this case it is also required that the absorbing gas is replenished. As bright phases of \xu12\ with luminosities above \oergs{39} are observed for more than 14 years, the source of the gas must be long-lasting. As the gas has to be rich in O, it must be processed material containing few H and He, like white dwarfs. It has been suggested that an RCB star can be the source of O rich gas \citep{2011MNRAS.410L..32M}, but such a star ejects parts of its envelope on time scales of years \citep{2012JAVSO..40..539C}, and hence would not be able to replenish the gas on the time scales observed in \xu12. Other possibilities to cause excess absorption are a grazing eclipse by the donor star, a disc wind from the accretion disc, or a grazing eclipse by a precessing, warped accretion disc \citep{2008MNRAS.386.2075S}. An eclipse by the donor star cannot explain the change in the X-ray flux \citep{2008MNRAS.386.2075S}. The same authors derived an about 90 d precession period for the warped disc. In other systems such as Her X--1, SS433 and LMC X--4 the super-orbital period is several tens to hundred of days \citep{2011MNRAS.418..437J}. Therefore, the observed variability in \xu12 is inconsistent with the time scales expected for a precessing, warped accretion disc. As disc winds occur on aperiodic time scales, a disc wind from a compact object accreting above Eddington remains a viable explanation for the light curve of \xu12. The radius of the thermal emission area and its temperature seem
consistent with hyper-Eddington black hole winds \citep[see][]{2016MNRAS.455.1211K}.


\acknowledgments
This work is based on observations obtained with \xmm, an ESA science mission with instruments and contributions directly funded by ESA Member States and NASA. Data used in this work are available from the \xmm\ Science Archive (http://xmm.esac.esa.int/xsa/). This project is supported by the Ministry of Science and Technology of the Republic of China (Taiwan) through grants 105-2112-M-007-033-MY2 and 105-2811-M-007-065.

{\it Facilities:} \facility{\xmm}.

\bibliographystyle{/Users/holger/work/papers/maxi1535/apj}

\begin{thebibliography}{}
\expandafter\ifx\csname natexlab\endcsname\relax\def\natexlab#1{#1}\fi

\bibitem[{{Allen} {et~al.}(2011){Allen}, {Hewett}, {Maddox}, {Richards}, \&
  {Belokurov}}]{2011MNRAS.410..860A}
{Allen}, J.~T., {Hewett}, P.~C., {Maddox}, N., {Richards}, G.~T., \&
  {Belokurov}, V. 2011, \mnras, 410, 860

\bibitem[{{Bachetti} {et~al.}(2014){Bachetti}, {Harrison}, {Walton},
  {Grefenstette}, {Chakrabarty}, {F{\"u}rst}, {Barret}, {Beloborodov}, {Boggs},
  {Christensen}, {Craig}, {Fabian}, {Hailey}, {Hornschemeier}, {Kaspi},
  {Kulkarni}, {Maccarone}, {Miller}, {Rana}, {Stern}, {Tendulkar}, {Tomsick},
  {Webb}, \& {Zhang}}]{2014Natur.514..202B}
{Bachetti}, M., {Harrison}, F.~A., {Walton}, D.~J., {et~al.} 2014, \nat, 514,
  202

\bibitem[{{Clayton}(2012)}]{2012JAVSO..40..539C}
{Clayton}, G.~C. 2012, Journal of the American Association of Variable Star
  Observers (JAAVSO), 40, 539

\bibitem[{{Dickey} \& {Lockman}(1990)}]{1990ARA&A..28..215D}
{Dickey}, J.~M., \& {Lockman}, F.~J. 1990, \araa, 28, 215

\bibitem[{{Evans} \& {Kochanek}(1989)}]{1989ApJ...346L..13E}
{Evans}, C.~R., \& {Kochanek}, C.~S. 1989, \apjl, 346, L13

\bibitem[{{Golenetskii} {et~al.}(2003){Golenetskii}, {Aptekar}, {Frederiks},
  {Mazets}, {Palshin}, {Hurley}, {Cline}, \& {Stern}}]{2003ApJ...596.1113G}
{Golenetskii}, S., {Aptekar}, R., {Frederiks}, D., {et~al.} 2003, \apj, 596,
  1113

\bibitem[{{Guillochon} \& {Ramirez-Ruiz}(2013)}]{2013ApJ...767...25G}
{Guillochon}, J., \& {Ramirez-Ruiz}, E. 2013, \apj, 767, 25

\bibitem[{{Houck} \& {Denicola}(2000)}]{2000ASPC..216..591H}
{Houck}, J.~C., \& {Denicola}, L.~A. 2000, in Astronomical Society of the
  Pacific Conference Series, Vol. 216, Astronomical Data Analysis Software and
  Systems IX, ed. {N.~Manset, C.~Veillet, \& D.~Crabtree}, 591

\bibitem[{{Irwin} {et~al.}(2016){Irwin}, {Maksym}, {Sivakoff}, {Romanowsky},
  {Lin}, {Speegle}, {Prado}, {Mildebrath}, {Strader}, {Liu}, \&
  {Miller}}]{2016Natur.538..356I}
{Irwin}, J.~A., {Maksym}, W.~P., {Sivakoff}, G.~R., {et~al.} 2016, \nat, 538,
  356

\bibitem[{{Joseph} {et~al.}(2015){Joseph}, {Maccarone}, {Kraft}, \&
  {Sivakoff}}]{2015MNRAS.447.1460J}
{Joseph}, T.~D., {Maccarone}, T.~J., {Kraft}, R.~P., \& {Sivakoff}, G.~R. 2015,
  \mnras, 447, 1460

\bibitem[{{Jurua} {et~al.}(2011){Jurua}, {Charles}, {Still}, \&
  {Meintjes}}]{2011MNRAS.418..437J}
{Jurua}, E., {Charles}, P.~A., {Still}, M., \& {Meintjes}, P.~J. 2011, \mnras,
  418, 437

\bibitem[{{Keek} {et~al.}(2006){Keek}, {in't Zand}, \&
  {Cumming}}]{2006A&A...455.1031K}
{Keek}, L., {in't Zand}, J.~J.~M., \& {Cumming}, A. 2006, \aap, 455, 1031

\bibitem[{{King} \& {Muldrew}(2016)}]{2016MNRAS.455.1211K}
{King}, A., \& {Muldrew}, S.~I. 2016, \mnras, 455, 1211

\bibitem[{{Kuulkers}(2004)}]{2004NuPhS.132..466K}
{Kuulkers}, E. 2004, Nuclear Physics B Proceedings Supplements, 132, 466

\bibitem[{{Maccarone} {et~al.}(2007){Maccarone}, {Kundu}, {Zepf}, \&
  {Rhode}}]{2007Natur.445..183M}
{Maccarone}, T.~J., {Kundu}, A., {Zepf}, S.~E., \& {Rhode}, K.~L. 2007, \nat,
  445, 183

\bibitem[{{Maccarone} {et~al.}(2014){Maccarone}, {Lehmer}, {Leyder},
  {Antoniou}, {Hornschemeier}, {Ptak}, {Wik}, \& {Zezas}}]{2014MNRAS.439.3064M}
{Maccarone}, T.~J., {Lehmer}, B.~D., {Leyder}, J.~C., {et~al.} 2014, \mnras,
  439, 3064

\bibitem[{{Maccarone} \& {Warner}(2011)}]{2011MNRAS.410L..32M}
{Maccarone}, T.~J., \& {Warner}, B. 2011, \mnras, 410, L32

\bibitem[{{Neronov} \& {Vovk}(2011)}]{2011MNRAS.412.1389N}
{Neronov}, A., \& {Vovk}, I. 2011, \mnras, 412, 1389

\bibitem[{{Phinney}(1989)}]{1989IAUS..136..543P}
{Phinney}, E.~S. 1989, in IAU Symposium, Vol. 136, The Center of the Galaxy,
  ed. M.~{Morris}, 543

\bibitem[{{Rees}(1988)}]{1988Natur.333..523R}
{Rees}, M.~J. 1988, \nat, 333, 523

\bibitem[{{Rhode} \& {Zepf}(2001)}]{2001AJ....121..210R}
{Rhode}, K.~L., \& {Zepf}, S.~E. 2001, \aj, 121, 210

\bibitem[{{Shih} {et~al.}(2008){Shih}, {Maccarone}, {Kundu}, \&
  {Zepf}}]{2008MNRAS.386.2075S}
{Shih}, I.~C., {Maccarone}, T.~J., {Kundu}, A., \& {Zepf}, S.~E. 2008, \mnras,
  386, 2075

\bibitem[{{Sivakoff} {et~al.}(2005){Sivakoff}, {Sarazin}, \&
  {Jord{\'a}n}}]{2005ApJ...624L..17S}
{Sivakoff}, G.~R., {Sarazin}, C.~L., \& {Jord{\'a}n}, A. 2005, \apjl, 624, L17

\bibitem[{{Steele} {et~al.}(2011){Steele}, {Zepf}, {Kundu}, {Maccarone},
  {Rhode}, \& {Salzer}}]{2011ApJ...739...95S}
{Steele}, M.~M., {Zepf}, S.~E., {Kundu}, A., {et~al.} 2011, \apj, 739, 95

\bibitem[{{Steiner} {et~al.}(2009){Steiner}, {Narayan}, {McClintock}, \&
  {Ebisawa}}]{2009PASP..121.1279S}
{Steiner}, J.~F., {Narayan}, R., {McClintock}, J.~E., \& {Ebisawa}, K. 2009,
  \pasp, 121, 1279

\bibitem[{{Strohmayer} \& {Bildsten}(2003)}]{2003astro.ph..1544S}
{Strohmayer}, T., \& {Bildsten}, L. 2003, ArXiv Astrophysics e-prints,
  astro-ph/0301544

\bibitem[{{Tsygankov} {et~al.}(2017){Tsygankov}, {Doroshenko}, {Lutovinov},
  {Mushtukov}, \& {Poutanen}}]{2017arXiv170200966T}
{Tsygankov}, S.~S., {Doroshenko}, V., {Lutovinov}, A.~A., {Mushtukov}, A.~A.,
  \& {Poutanen}, J. 2017, ArXiv e-prints, arXiv:1702.00966

\bibitem[{{Tully} {et~al.}(2008){Tully}, {Shaya}, {Karachentsev}, {Courtois},
  {Kocevski}, {Rizzi}, \& {Peel}}]{2008ApJ...676..184T}
{Tully}, R.~B., {Shaya}, E.~J., {Karachentsev}, I.~D., {et~al.} 2008, \apj,
  676, 184

\bibitem[{{Vagnetti} {et~al.}(2016){Vagnetti}, {Middei}, {Antonucci},
  {Paolillo}, \& {Serafinelli}}]{2016A&A...593A..55V}
{Vagnetti}, F., {Middei}, R., {Antonucci}, M., {Paolillo}, M., \&
  {Serafinelli}, R. 2016, \aap, 593, A55

\bibitem[{{Zdziarski} {et~al.}(1996){Zdziarski}, {Johnson}, \&
  {Magdziarz}}]{1996MNRAS.283..193Z}
{Zdziarski}, A.~A., {Johnson}, W.~N., \& {Magdziarz}, P. 1996, \mnras, 283, 193

\bibitem[{{Zepf} {et~al.}(2008){Zepf}, {Stern}, {Maccarone}, {Kundu},
  {Kamionkowski}, {Rhode}, {Salzer}, {Ciardullo}, \&
  {Gronwall}}]{2008ApJ...683L.139Z}
{Zepf}, S.~E., {Stern}, D., {Maccarone}, T.~J., {et~al.} 2008, \apjl, 683, L139

\bibitem[{{{\.Z}ycki} {et~al.}(1999){{\.Z}ycki}, {Done}, \&
  {Smith}}]{1999MNRAS.309..561Z}
{{\.Z}ycki}, P.~T., {Done}, C., \& {Smith}, D.~A. 1999, \mnras, 309, 561

\end{thebibliography}




\end{document}